# Asymmetric Tsallis distributions for modelling financial market dynamics

Sandhya Devi[†]

509 6[th] Ave S, Edmonds, WA 98020, United States of America

Email: sdevi@entropicdynamics.com

**Abstract:** Financial markets are highly non-linear and non-equilibrium systems. Earlier works have suggested that the behavior of market returns can be well described within the framework of non-extensive Tsallis statistics or superstatistics. For small time scales (delays), a good fit to the distributions of stock returns is obtained with $q$-Gaussian distributions, which can be derived either from Tsallis statistics or superstatistics. These distributions are symmetric. However, as the time lag increases, the distributions become increasingly non-symmetric. In this work, we address this problem by considering the data distribution as a linear combination of two independent normalized distributions – one for negative returns and one for positive returns. Each of these two independent distributions are half $q$-Gaussians with different non-extensivity parameter $q$ and temperature parameter beta. Using this model, we investigate the behavior of stock market returns over time scales from 1 to 80 days. The data covers both the .com bubble and the 2008 crash periods. These investigations show that for all the time lags, the fits to the data distributions are better using asymmetric distributions than symmetric $q$-Gaussian distributions. The behaviors of the $q$ parameter are quite different for positive and negative returns. For positive returns, $q$ approaches a constant value of 1 after a certain lag, indicating the distributions have reached equilibrium. On the other hand, for negative returns, the $q$ values do not reach a stationary value over the time scales studied. In the present model, the markets show a transition from normal to superdiffusive behavior (a possible phase transition) during the 2008 crash period. Such behavior is not observed with a symmetric $q$-Gaussian distribution model with $q$ independent of time lag.

## 1. Introduction

Many well-known financial models [1] are based on the efficient market hypothesis [2] according to which: a) investors have all the information available to them and they independently make rational decisions using this information, b) the market reacts to all the information available reaching equilibrium quickly, and c) in this equilibrium state the market essentially follows a random walk [3]. In such a system, extreme changes are very rare. In reality however, the market is a complex system that is the result of decisions by interacting agents (e.g., herding behavior), traders who speculate and/or act impulsively on little news, etc. Such collective/chaotic behavior can lead to wild swings in the system, driving it away from equilibrium into the realm of nonlinearity, resulting in a variety of interesting phenomena such as phase transitions, critical phenomena such as bubbles, crashes [4], superdiffusion [5] and so on.

---

[†] Shell International Exploration and Production Co. (Retired)

The distribution of a system following a random walk can be obtained by maximizing Shannon entropy [6][7] with constraints on the moments. In particular, this maximization with constraints on the normalization and the second moment yields a Gaussian distribution. Therefore, if the stock market follows a pattern of random walk, the corresponding returns should show a Gaussian distribution. However, it is well known [8] that stock market returns, in general, have a more complicated distribution. This is illustrated in figure 1 which compares the distributions of 1 day and 20 day log returns of S&P 500 and Nasdaq stock markets (1994-2014) with the corresponding Gaussian distributions. The data distributions show sharp peaks in the center and fat tails over many scales, neither of which is captured by the Gaussian distribution. Several studies [9][10] indicate that these issues can be addressed using statistical methods based on Tsallis entropy [11] (also known as $q$-entropy), which is a generalization of Shannon entropy to non-extensive systems. These formalisms were originally proposed to study classical and quantum chaos, physical systems far from equilibrium such as turbulent systems, and long range interacting Hamiltonian systems. However, in the last several years, there has been considerable interest in applying these methods to analyze financial market dynamics as well. Such applications fall into the category of econophysics [5].

As in the case of Shannon entropy, the equilibrium distribution of a non-extensive system can be obtained by maximizing Tsallis entropy with constraints on the moments of the distribution [11]. Constraints on the normalization and the second moment yield a symmetric $q$-Gaussian distribution. As $q \to 1$, the system tends to an extensive system and the $q$-Gaussian distribution approaches a Gaussian distribution.

Let us now look at the non-equilibrium situation. In the random walk model, the dynamics of the probabilities are described by a Fokker Plank (FP) equation with linear drift and diffusion terms. The solution to this FP equation is a Gaussian distribution. In this case, the autocorrelation of the standard deviation $\sigma$ falls as $\sqrt{\tau}$ where $\tau$ is the time delay. A generalization of Fokker-Plank equation to non-extensive systems, using statistical methods based on Tsallis entropy, has been given by [12][13]. This results in a Fokker-Plank equation with a linear drift term and a non-linear diffusion term. The solution to this equation, under some assumptions, is a Tsallis $q$-Gaussian distribution, with the parameter $q$ independent of time. For $q < 5/3$, the generalized inverse mean square deviation of this distribution follows a power law, with the magnitude of the exponent > 1. This points to a superdiffusive process. This model has been applied to study both high frequency [14][15] and long-term low frequency stock market returns [16]. However, in a previous publication [17] we have shown that for low frequency returns (time delay >= 1 day) with $q$ independent of time delay, the inverse mean square deviation shows a power law behavior with the exponent very close to 1 pointing to a normal diffusive behavior. Hence, for returns longer than one day, the predicted superdiffusive behavior from the non-linear FP equations discussed above is not supported by the empirical results.

The non-extensive statistics is relevant for two classes of systems:

a) Systems which have interacting sub-systems: Tsallis formulation of non-extensivity based on entropy principle is used for this class. In this case, the entropies of the sub-systems are not additive.

b) Superstatistics: Beck [18] showed that non-extensivity can also arise in systems following ordinary statistics if the temperature or energy dissipation rate is locally fluctuating. In this case, the distribution can be considered as the integral of the probability density $f(\beta)$ of the inverse temperature $\beta$ with the Boltzmann-Gibbs (BG) distribution function $e^{-\beta E}$. This is a superposition of two statistics – hence the name 'superstatistics'. When $\beta$ has a $\chi^2$ distribution (which is widely applicable to many common systems [19]), the superstatistics yields a $q$-exponential or $q$-Gaussian distribution, depending on the definition of E [20][21].

One thing to note in the case of superstatistics is that even though the system follows a BG distribution, because of the temperature variation, the entropy is quasi-additive [18]. This means, if we consider two independent systems $A$ and $B$ having some value $q$, the entropy of the composite system $A + B$ is the sum of the entropies of $A$ and $B$ but with a different $q$. For dynamical systems, if $A$ and $B$ belong to smaller time scales, the system with a larger time scale will have a different $q$. Hence in superstatistics, the non-extensivity parameter changes with time scale. We will discuss the dynamics of $q$ in more detail in section 2 since this is an important aspect of the present analysis.

Neither of the two non-extensivity models discussed above supports asymmetry. A perturbative approach to treating asymmetry has been discussed in [22]. However, a look at figure 2, which displays the variation of skew with time lag for S&P 500 and Nasdaq returns for the .com bubble and the 2008 crash periods, clearly shows that the skew rises very rapidly with time lag. This is particularly pronounced during the crash period. Hence, except for very small time scales, the perturbative approach for treating asymmetry will not be adequate.

Budini [23] has derived a family of extended $q$-Gaussian distributions starting from two gamma distributions that have different shape parameters but the same temperature parameters. The resulting distribution is asymmetric only in the shape parameter and not in the spatial scale parameter.

In this paper we address the asymmetry in the distribution of financial market returns by requiring that both $q$ and the 'temperature' parameter $\beta$ depend on the sign of the market returns. Since asymmetry increases with time delay, both these parameters change with time scale.

The rest of paper is organized as follows. A brief review of derivations of $q$-Gaussian distributions from Tsallis entropic formulation and superstatistics will be given in section 2. Our formulation of asymmetric distributions will be discussed in section 3. Results of application of this formulation to financial market dynamics will be given in section 4 followed by summary and conclusions in section 5.

## 2. *q*-Gaussian distributions

### 2.1 Tsallis statistics

A detailed review of the derivation of $q$-Gaussian distribution from Tsallis entropy and the estimation of the parameters of the model is given in an earlier publication [17]. As already discussed earlier, in Tsallis statistics the non-extensivity of a system arises due to the long range interactions of the sub-systems. This is evident from the non-additive property of Tsallis entropy

$$S_q(A+B) = S_q(A) + S_q(B) + (1-q)\,S_q(A)\,S_q(B) \tag{1}$$

The third term on the right hand side of the equation (1) is the result of interaction between the sub-systems $A$ and $B$. As the non-extensivity parameter $q \to 1$, the additive property of the entropies is recovered. Note that $q$ for the combined system is the same as for the individual systems. An important thing to note here is that if $A$ and $B$ refer to financial market returns at two shorter time scales, the bigger system $A + B$ referring to larger time lag has the same non-extensivity index $q$. So, in this case, $q$ is static with respect to time scale.

### 2.2 Superstatistics

Beck [18] showed that non-extensivity can also arise in systems following ordinary statistics, if the temperature or energy dissipation rate is locally fluctuating. In this case, the distribution $P$ can be considered as the integral of the distribution of the inverse temperature $\beta$ with the Boltzmann-Gibbs (BG) distribution function $e^{-\beta E}$.

$$P(E) = \left(\frac{1}{N}\right) \int_0^\infty f(\beta) e^{-\beta E} d\beta \tag{2}$$

Here $N$ is the normalization constant. The right hand side of equation (2) is the weighted average of a BG distribution function. When $\beta$ is chosen to have a $\chi^2$ distribution, the probability density function is

$$f(\beta) = \left(\Gamma(\alpha)\right)^{-1} \left(\frac{\alpha}{\beta_0}\right)^\alpha \beta^{(\alpha-1)} e^{(-\alpha\beta/\beta_0)} \tag{3}$$

where

$$\beta_0 = \int \beta f(\beta) d\beta$$

In this case, the integral in equation (2) can be shown to be [20][21] the $q$-exponential function

$$[1 + (q-1)\beta_0 E]^{1/(1-q)} \tag{4}$$

Here $1/(q-1) = \alpha$.

The integral in equation (2) holds for any positive E. In our case, we choose E to be the square of the stock market return $\Omega$. In this case equation (4) becomes the $q$-Gaussian function

$$[1 + (q-1)\beta_0\Omega^2]^{1/(1-q)} \tag{5}$$

The $q$-exponential function given in equation (4) is a special case of superstatistics where $\beta$ is chosen to have a $\chi^2$ distribution. There are many other distribution functions besides $\chi^2$ which can multiply the BG function in superstatistics. Some of these are discussed in [24][25][26]. For low frequency stock returns (one day or longer), a $\chi^2$ distribution is shown to be appropriate [26][27]. It should be noted that for superstatistics to be a valid model, the variation of $\beta$ should be slow compared to that of $\Omega$. This is shown to be in general true if there are no outliers like isolated sharp spikes in the data [25][26].

For $q > 1$, $q$ can now be given a physical meaning as the relative variance of $\beta$

$$(q-1) = [< \beta^2 > - < \beta >^2]/< \beta >^2 \tag{6}$$

Here the expectation value $< >$ is taken with respect to $f(\beta)$.

Let us now consider the non-equilibrium case. Note that unlike in Tsallis statistics, here there is no long-range interaction between sub-systems. However, the variation of $\beta$ gives rise to a quasi-additive property of the entropies [18]

$$S_q(A) + S_q(B) = S_{q'}(A+B) \tag{7}$$

If we consider $A$ and $B$ as systems with smaller time scales, then the one with larger time delay $A + B$ will have a different $q'$. In general, one can expect the temperature fluctuations of the smaller systems to be larger than those of the composite system and hence the $q$ to be a decreasing function of time scale. For $q$ close to 1, it can be shown [18] that $q'$ and $q$ are related as

$$(q'-1)/(q-1) = [1 + (< \ln P_i >)^2/< (\ln P_i)^2 >]^{-1} \tag{8}$$

Equation (8) shows that for $q$ values close to 1, $q$ monotonically decreases as the time lag increases. The $q$ variation with time lag is one of the important things that will be discussed in the present analysis.

## 3. Asymmetric $q$-Gaussian distributions

To generalize Tsallis statistics and superstatistics to include asymmetry, we consider a distribution (PDF) which is a linear combination of two independent normalized distributions $P_1(\Omega)$ and $P_2(\Omega)$

$$P(\Omega) = \frac{1}{2}\left[P_1(\Omega) + P_2(\Omega)\right] \tag{9}$$

where $\Omega = (x - \mu)/\sigma_0$ is a zero mean random variable, $\mu$ is the mean of the data variable $x$ and $\sigma_0$ is a standard deviation at the lowest time scale. Denoting the distributions for negative and positive returns $\Omega$ as $P_-$ and $P_+$ respectively, we rewrite (9) as

$$P(\Omega) = \left[P_-(\Omega) + P_+(\Omega)\right] \tag{10}$$

where

$$P_-(\Omega) = 0 \qquad \Omega > 0 \tag{11a}$$

$$P_+(\Omega) = 0 \qquad \Omega \leq 0 \tag{11b}$$

$P_-(\Omega)$ and $P_+(\Omega)$ are half $q$-Gaussians, given by

$$P_-(\Omega) = \frac{1}{Z_-}\left[1 + (q_- - 1)\beta_-\Omega^2\right]^{1/(1-q_-)} \qquad \Omega \leq 0 \tag{12a}$$

$$P_+(\Omega) = \frac{1}{Z_+}\left[1 + (q_+ - 1)\beta_+\Omega^2\right]^{1/(1-q_+)} \qquad \Omega > 0 \tag{12b}$$

Here, $q_-$, $\beta_-$, $q_+$ and $\beta_+$ are the $q$-Gaussian parameters for negative and positive $\Omega$ respectively. The normalizations $Z_-$ and $Z_+$ are such that

$$\int_{-\infty}^{0} P_-(\Omega)\,d\Omega = \int_{0}^{\infty} P_+(\Omega)\,d\Omega = 1/2 \tag{13}$$

so that the complete PDF $P(\Omega)$ in (10) is normalized. This gives

$$Z_- = C_{q_-}/(\sqrt{\beta_-}) \tag{14a}$$

$$Z_+ = C_{q_+}/(\sqrt{\beta_+}) \tag{14b}$$

$$C_q = \sqrt{\pi}\,\frac{\Gamma\left(\frac{1}{q-1} - \frac{1}{2}\right)}{\sqrt{q-1}\,\Gamma\left(\frac{1}{q-1}\right)} \tag{14c}$$

Equations (11a) – (14c) can be derived from the maximization of $q$-entropy assuming $q$ is dependent on the sign of $\Omega$ and giving separate constraints to negative and positive $\Omega$ (see appendix A). In this case, $\beta_-$ and $\beta_+$ are the Lagrange multipliers for the constraints on the variance. The same results can be obtained from superstatistics by letting the parameters $\alpha$ and $\beta_0$ in $f(\beta)$ be dependent on the sign of $\Omega$. In this case $\beta_-$ and $\beta_+$ can be interpreted as the expectation values of the temperature with respect to the temperature distributions for the negative and positive values of the variable $\Omega$.

Since $\Omega$ has zero mean, the mean square deviation for the asymmetric case is given by

$$\sigma_{asym}^2 = \int \Omega^2 P(\Omega) d\Omega$$

$$= \frac{1}{2} \left[ 1/(\beta_-(5 - 3q_-)) + 1/(\beta_+(5 - 3q_+)) \right] \tag{15}$$

for $q_-$, $q_+ < 5/3$.

Equations (9) – (15) go over to symmetric case results when $q_- = q_+ = q$ and $\beta_- = \beta_+ = \beta$.

Let us now consider the non-equilibrium case. As discussed in section 2, in the Tsallis entropic model the parameter $q$ is independent of time scale. However, in the superstatistics model $q$ varies with time scale. Equation (8) gives the variation for small $(q - 1)$. Since $P_-$ and $P_+$ are independent, the parameters $q_-$ and $q_+$ are also independent. Hence following a procedure similar to that given in [18] one can show (see appendix B) that an equivalent relationship can be obtained for the asymmetric case.

$$(q_-' - 1)/(q_- - 1) = [1 + (< \ln P_- >)^2/< (\ln P_-)^2 >]^{-1} \tag{16a}$$

$$(q_+' - 1)/(q_+ - 1) = [1 + (< \ln P_+ >)^2/< (\ln P_+)^2 >]^{-1} \tag{16b}$$

The expectation values $< >$ are taken with the corresponding distributions. Note that the $q$ for each branch monotonically decrease with time scale.

## 4. Results

The data chosen for our analysis are S&P 500 and Nasdaq daily (close of the day) stock prices which are de-trended with CPI to remove inflation trends. These are displayed in figure 3. We will consider the period after 1991 (about a year before the time when electronic trading over the internet was launched), since the character of the stock price variation changes dramatically after that. The time series shows a non-stationary character with wild fluctuations. The data for analysis is divided into two regions bounded by vertical dotted lines. Regions 1 and 2 cover the dot-com bubble period and the crash of 2008 respectively. Each region has 3000 samples.

The variables used for the estimation of $q$ and $\beta$ are the standardized log returns $\Omega(t, t_0)$ for delay $t$:

$$\Omega(t, t_0) = (y(t, t_0) - \mu_t)/\sigma_1 \tag{17}$$

computed for several starting times $t_0$ over the period of interest. Here

$$y(t, t_0) = \ln X(t_0 + t) - \ln X(t_0)$$

$X$ is the stock value, $\mu_t$ is the mean of $y(t)$, and $\sigma_1$ is the standard deviation for 1 day log returns. With this choice, $\bar{\Omega} = 0$. The 1 day standardized log returns are displayed in figure 4. Note that in general the negative log returns show relatively more higher spikes than the positive returns.

In [17], we discussed the estimation of the $q$-Gaussian parameters for the symmetric case using the maximum likelihood estimate method [28]. Here the $q$ parameter was held constant as a function of time scale. We follow a similar procedure here to estimate the parameters $q_-$, $\beta_-$ and $q_+$, $\beta_+$ from data $\Omega \leq 0$ and $\Omega > 0$ respectively, for time delays 1-80 days. This range is long enough to study the asymmetric effects and short enough so that the number of samples for the parameter estimation is not drastically reduced.

## 4.1 Goodness of fit

Figures 5 and 6 show the comparison of the asymmetric $q$-Gaussian distributions obtained from the estimated parameters, with the data distributions for regions 1 and 2 respectively. All the PDF's have large tails particularly for negative returns. This is the result of large jumps in the time series of returns shown in figure 4. One thing to note is that even at large time delays, the distributions do not approach Gaussian distributions (central limit theorem). In fact, the asymmetry increases with delay.

To quantify the goodness of the parameter estimates, Kolmogorov-Smirnov (KS) tests [29] are performed at all delays considered. To do this, synthetic data are generated at each delay using a generalized Box-Müller method for generating $q$-Gaussian random deviates [30] given $q$, $\beta$ values. The synthetic data are standardized in the same way as the empirical data. The maximum absolute distances $\mathbf{D_{max}}$ between the empirical and synthetic cumulative distribution functions (CDF) are calculated. If $\mathbf{D_{max}}$ exceeds a critical distance $\mathbf{D_{crit}}$ [31] at a particular significance level, that fit should either be rejected or accepted at a higher significance level. $\mathbf{D_{crit}}$ is given by

$$\mathbf{D_{crit}} = c(\gamma) \sqrt{(n1 + n2)/(n1 * n2)}$$

Here, *n1* and *n2* are the number of samples in the empirical and synthetic CDF's respectively. The table for function $c(\gamma)$ at different significance levels $\gamma$ can be found in [31].

Figure 7 shows $\mathbf{D_{max}}$ as a function of delay for the negative and positive returns branches. Also shown are the critical distances $\mathbf{D_{crit}}$ for a significance level of 0.05 (confidence 95%). The distances in general are all well below the corresponding $\mathbf{D_{crit}}$. However, for region 2 (the crash period of 2008) negative returns, $\mathbf{D_{max}}$ increases for higher delays. A look at figure 8, which shows the S&P 500 70 day log returns, explains why this is so. There are a few big spikes for negative returns which dominate the estimates resulting in higher errors. But even for these delays, $\mathbf{D_{max}}$ is well below the critical level.

For comparison, the KS test results for the symmetric $q$-Gaussian fit are displayed in figure 9. For easier comparison with the asymmetric case, the results for negative and positive returns are shown separately. In almost all cases, the $\mathbf{D_{max}}$ is much higher and closer to $\mathbf{D_{crit}}$ than in the asymmetric case. Hence any conclusions drawn on the behavior of the estimated parameters are questionable. This is particularly so for the negative returns indicating the importance of considering asymmetry.

## 4.2 Parameters of the asymmetric Tsallis distribution as functions of time scale

The asymmetric character of the distributions become more obvious when we look at the variation of $q_-$ and $q_+$ with time scale (figure 10). For positive returns, $q_+$ is greater than 1 but less than the limit 5/3 (equation (15)) for smaller delays (< 20 days) and approaches the value of 1 after that. The attractor in this case is a (half) Gaussian which has a finite variance. The behavior is quite different for negative returns. No stationary state is reached even at 80 days. For region 2, the $q_-$ values at longer delays are higher than the limit 5/3 and continue to slowly increase. The high values of $q_-$ are the results of very fat tails as shown in figure 6. Note that, even though the variance is infinite for $q > 5/3$, the attractor (steady state distribution) in this case could be an $\alpha$-stable distribution which has infinite variance and satisfies a generalized central limit theorem [32].

The variation of the estimated $\beta_-$ and $\beta_+$, along with error bars, are shown in figure 11 on a log-log scale. The two plots are very close to each other indicating the main contribution to asymmetry comes from $q$ and not from the temperature parameter. The straight line character of the plots shows that $\beta$ for both negative and positive returns follows a power law behavior.

## 4.3 Superdiffusion

In an earlier publication [17] we showed that in the framework of Tsallis statistics, where the entropic index $q$ is independent of time scale, no superdiffusion behavior is observed either during the .com bubble or the 2008 crash periods. The variance estimated from the symmetric $q$-Gaussian distribution increases almost linearly with time scale. This points to a normal behavior.

The case for asymmetric distributions is given in figure 12, which shows the variation of the mean square deviation $\sigma^2_{asym}$ (equation (15)) with time delay, on a log-log scale. The straight line character of the plots shows that

$$\sigma^2_{asym} \propto t^{\lambda}$$

In region 1, the $\lambda$ values are close to 1 indicating normal diffusion. A value of $\lambda$ slightly greater than 1 for Nasdaq (which was greatly affected by the .com bubble) indicates a mild superdiffusive behavior. In region 2, $\lambda$ is $> 1$. This clearly indicates a superdiffusive behavior. At larger time delays $q_-$ exceeds the limit of 5/3 and $\sigma^2_{asym}$ is no longer defined. These delays are omitted from figure 12. This transition from normal diffusion in region 1 to superdiffusion in region 2 indicates a possible phase transition.

### 4.4 Superstatistics

Beck showed [18] that when $(q - 1)$ is small, $1/(q - 1) \propto r^{\delta}$ where $r$ is the spatial scale. Laboratory observations of hydrodynamic turbulence give a value of ~ 0.3 for $\delta$. Theoretical estimation by Beck gives a value of $\delta \sim 0.4$. Figure 13 shows $1/(q - 1)$ versus time scale for delays up to 20 days, since $q_+ \to 1$ after this. For positive returns, $\delta$ lies between 0.33 - 0.55 which is close to both the laboratory observations and the theoretical estimates of Beck for hydrodynamic turbulence. However, the situation is quite different for negative returns. The $(q_- - 1)$ values do not decrease monotonically with delay. The values remain high at all times and hence the perturbation result (16a) may not hold. The high $q_-$ values are the results of much fatter tails on the negative return side than on the positive return side (figures 5 and 6). This might be an indication that during heavy market sell off, the reaction of agents is heavily dependent on what others are doing (herding behavior). This in turn indicates that the non-extensive character of the market in this case is likely arising due to long range interactions rather than from the local temperature variations.

## 5. Summary and Conclusions

Investigations of the behavior of the financial market long term returns, over a period which includes both the dot-com period of 2000 and the crash of 2008, show that the distributions of the returns are non-Gaussian and fat-tailed even for as long a term as 1-80 days. This behavior cannot be adequately described by Boltzmann-Gibbs statistics. One needs to consider non-extensive statistical mechanics such as Tsallis statistics and/or superstatistics.

Observations of the PDFs of market returns show that for small time scales, they can be well modelled by a symmetric Tsallis $q$-Gaussian distribution. However, as the time delay increases, the PDFs become increasingly asymmetric. In this work we generalize the Tsallis distribution to the asymmetric case by modelling the data distribution as a linear combination of two independent distributions each of which are half $q$-Gaussians and separately describe the PDF's corresponding to negative and positive returns.

The $q$-Gaussian parameters, namely the entropic index $q$ and the temperature $\beta$, are different for negative and positive returns and are estimated separately. These parameters are allowed to vary with time scale, since the asymmetry changes with time scale.

The goodness of fit tests (KS tests) show that the asymmetric model distributions fit very well the data distributions for all delays. The same tests performed on the corresponding symmetric model distributions show considerably higher errors.

The time scale behavior of the entropic indices $q_-$ and $q_+$ and the temperature parameters $\beta_-$ and $\beta_+$ show that the asymmetry arises mainly due to the quite different dynamical behaviors of the entropic indices for negative and positive market returns. In all cases, $q_+$ decreases monotonically with time delay approaching an asymptotic value of 1 in a relatively short period (20 days). The equilibrium distribution in this case is a Gaussian distribution. This behavior is consistent with the superstatistics model put forth by Beck. In contrast, the behavior of $q_-$ is very different. The value remains high at all times, even increasing slowly at higher delays. No stationary state is reached during the period of study. Heavy market sell offs are usually due to strong interactions between agents (herding behavior). Hence, in this case, the non-extensivity of the system might be best described by Tsallis statistics.

Finally, the system shows a phase transition from a normal diffusive state during the .com bubble period to a superdiffusive state during crash period. This is indicated by the power law behavior of the variance with the exponent changing from ~1 to a value > 1.

The present investigations show the importance of including asymmetry for an adequate description of stock market return distributions. The results show that both superstatistics and Tsallis statistics non-extensive models are needed to describe the complex financial market dynamics.

## Acknowledgements

Many thanks to Sherman Page for a critical reading of the manuscript.

## Appendix A:  Asymmetric $q$-Gaussian distributions from Tsallis statistics

The Tsallis $q$ entropy is given by

$$S = \sum_i P(\Omega_i)\, ln_q\big(1/P(\Omega_i)\big) \qquad\qquad (A1)$$

where $P(\Omega_i)$ is the probability density function at sample $i$ of the variable $\Omega$ and the $q$ logarithm $ln_q(x)$ is given by

$$ln_q(x) = (x^{1-q} - 1)/(1-q) \tag{A2}$$

In order to consider asymmetry, we make $q$ dependent on the sign of the $\Omega$. Using equations (10) and (A2), equation (A1) can be written as

$$S = \sum_{\Omega \leq 0} P_- \, ln_{q_-}(1/P_-) + \sum_{\Omega > 0} P_+ \, ln_{q_+}(1/P_+) \tag{A3}$$

The parameters $q_-$ and $q_+$ are estimated by maximizing $S$ under suitable constraints. Assuming $\Omega$ has zero mean and considering the continuous case for the random variable $\Omega$,

$$S = \int_{-\infty}^{0} P_-(\Omega) \, ln_{q_-}(1/P_-(\Omega)) d\Omega \; + \int_{0}^{\infty} P_+(\Omega) \, ln_{q_+}(1/P_+(\Omega)) d\Omega$$

$$= \frac{1}{(1-q_-)} \int_{-\infty}^{0} [P_-^{q_-} - P_-] \, d\Omega \; + \frac{1}{(1-q_+)} \int_{0}^{\infty} [P_+^{q_+} - P_+] \, d\Omega$$

It is straightforward to show that maximizing S with respect to $P_-$ and $P_+$ under the constraints

$$\int_{-\infty}^{0} P_-(\Omega) \, d\Omega \; = \int_{0}^{\infty} P_+(\Omega) \, d\Omega \; = 1/2 \tag{A4}$$

$$\int_{-\infty}^{0} \Omega^2 \, P_-^{q-}(\Omega) \, d\Omega \; = \sigma_{q_-}^2 \tag{A5}$$

$$\int_{0}^{\infty} \Omega^2 \, P_+^{q+}(\Omega) \, d\Omega \; = \sigma_{q_+}^2 \tag{A6}$$

gives

$$P_-(\Omega) = \frac{1}{Z_-} \left[ 1 + (q_- - 1)\beta_- \Omega^2 \right]^{1/(1-q_-)} \qquad \Omega \leq 0 \tag{A7}$$

$$P_+(\Omega) = \frac{1}{Z_+} \left[ 1 + (q_+ - 1)\beta_+ \Omega^2 \right]^{1/(1-q_+)} \qquad \Omega > 0 \tag{A8}$$

with

$$Z_- = C_{q_-}/(\sqrt{\beta_-})$$

$$Z_+ = C_{q_+}/(\sqrt{\beta_+})$$

$$C_q = \sqrt{\pi} \, \frac{\Gamma\left(\dfrac{1}{q-1} - \dfrac{1}{2}\right)}{\sqrt{q-1} \, \Gamma\left(\dfrac{1}{q-1}\right)}$$

Here $\beta_-$ and $\beta_+$ are the Lagrange multipliers of the constraints (A5) and (A6).

## Appendix B:  Time scale variation of $q$

In the superstatistics model, the entropies are quasi-additive given by equation (7)

$$S_{q'}(A + B) = S_q(A) + S_q(B) \tag{B1}$$

For asymmetric distributions, we denote

$$q' = \{q'_-\ \ q'_+\} \ \text{ and } \ q = \{q_-\ \ q_+\} \tag{B2}$$

In deriving the following equations, we follow a procedure similar to one given by Beck [18].

For two independent sub-systems $A$ and $B$ with identical PDF's

$$S_q(A) + S_q(B) = 2S_q$$

$$= 2 \, \frac{\Sigma_{i-}\left(P_{i-} - P_{i-}^{q_-}\right)}{q_- - 1} \ + \ 2 \, \frac{\Sigma_{i+}\left(P_{i_+} - P_{i_+}^{q_+}\right)}{q_+ - 1} \tag{B3}$$

Writing

$$P^q = P e^{(q-1)\ln P}$$

and expanding the exponential to the order $(q-1)^2$, it is straightforward to show

$$S_q(A) + S_q(B) = [-2 < \ln P_- > - (q_- - 1) < (\ln P_-)^2 >]$$

$$+ \ [-2 < \ln P_+ > - (q_+ - 1) < (\ln P_+)^2 >] \tag{B4}$$

Here the expectation values $<\ >$ are taken with the corresponding distributions.

Now let us consider the entropy of the joint system $A + B$.  The entropy in this case is given by

$$S_{q'}(A + B) = \frac{\sum_{k_-}\left(\mathbb{P}_{k_-} - \mathbb{P}_{k_-}^{q_-}\right)}{q'_- - 1} + \frac{\sum_{k_+}\left(\mathbb{P}_{k_+} - \mathbb{P}_{k_+}^{q_+}\right)}{q'_+ - 1} \tag{B5}$$

Here the joint probabilities are

$$\mathbb{P}_{k_-} = P_{i_- j_-} = P_{i_-} P_{j_-} \text{ and } \mathbb{P}_{k_+} = P_{i_+ j_+} = P_{i_+} P_{j_+} \tag{B6}$$

where the indices $i$, $j$ are the indices of the subsystems and the summation over joint probability index $k$ stands for summation over $i$, $j$. Using (B5) and (B6), it is straightforward to show that

$$S_{q'}(A + B) = -2 < \ln P_- >$$

$$-(q_- - 1)[(< \ln P_- >)^2 + < (\ln P_-)^2 >]$$

$$-2 < \ln P_+ >$$

$$-(q_+ - 1)[(< \ln P_+ >)^2 + < (\ln P_+)^2 >] \tag{B7}$$

Here again, we have used $P^q = P e^{(q-1)\ln P}$ and expanded the exponential up to the order $(q-1)^2$.

Equating (B4) and (B7) and noting $P_-$ and $P_+$ are independent, we get

$$(q'_- - 1)/(q_- - 1) = [1 + (< \ln P_- >)^2 / < (\ln P_-)^2 >]^{-1} \tag{B8a}$$

$$(q'_+ - 1)/(q_+ - 1) = [1 + (< \ln P_+ >)^2 / < (\ln P_+)^2 >]^{-1} \tag{B8b}$$

**Figures**

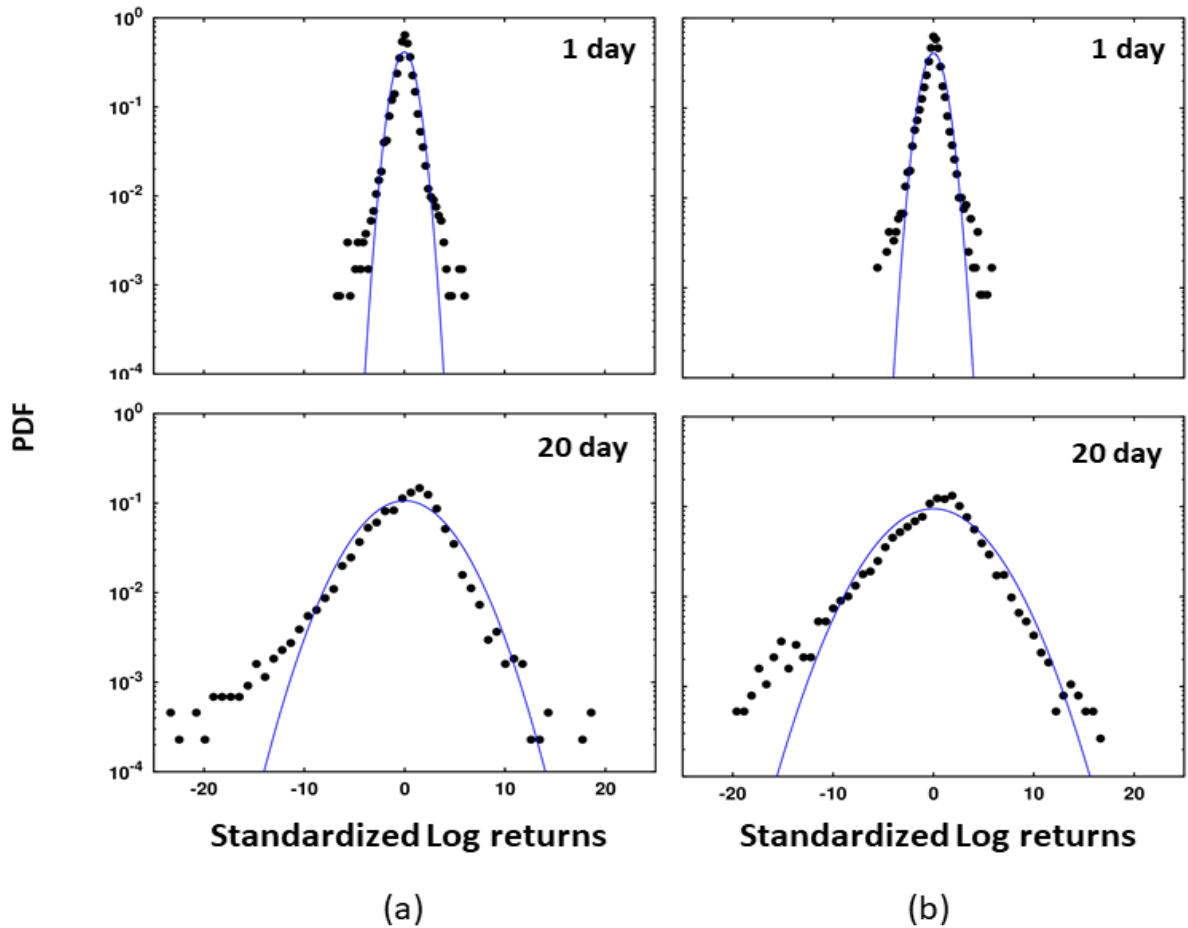

Figure 1. Comparison of the distributions of standardized log returns with the Gaussian distributions (solid blue line) having the same mean and standard deviation as the data (black dots). (a) S&P 500 for 2 January 1994 – 31 December 2013 and (b) Nasdaq over the same period.

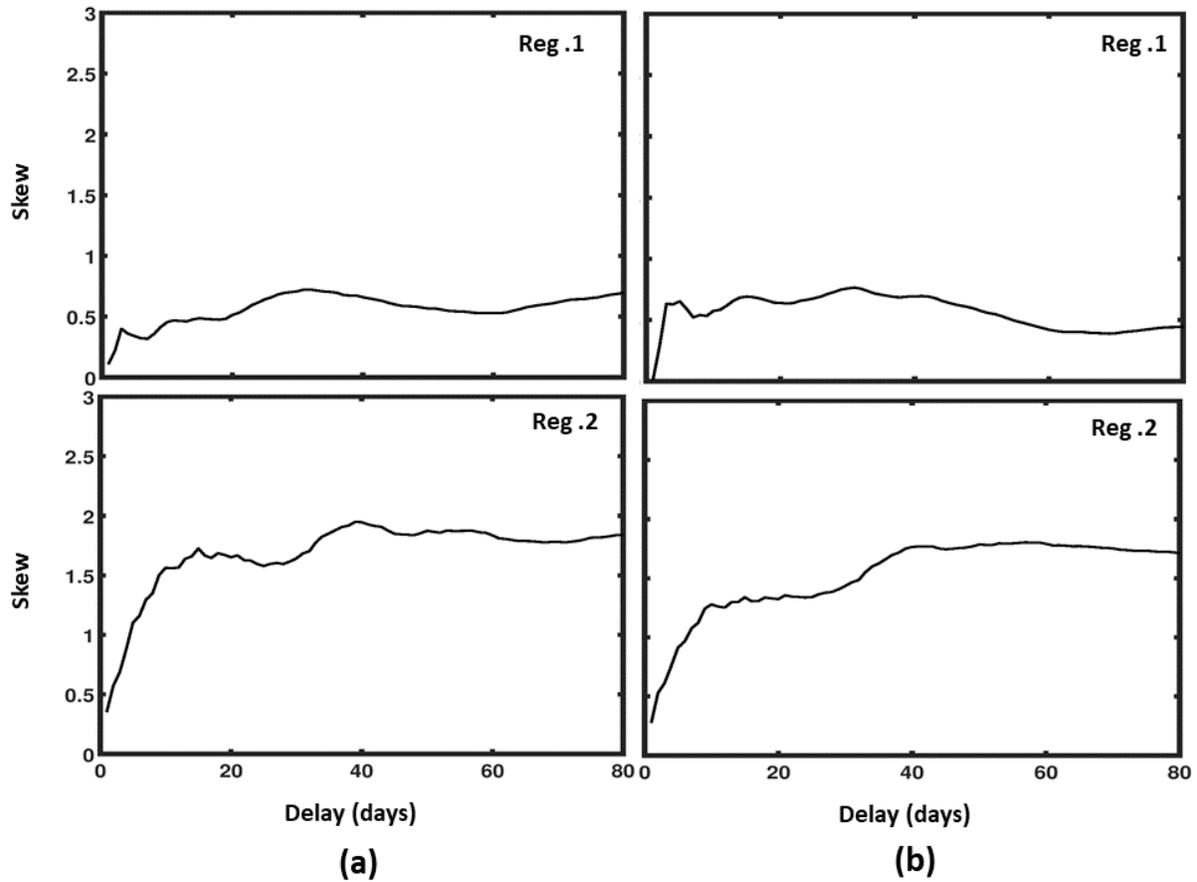

Figure 2. Skew of the PDF as a function of time scale (delay). Region 1 covers 14 December 1993 – 8 November 2005 and region 2 covers 9 November 2005 – 11 October 2017 for (a) S&P 500 and (b) Nasdaq.

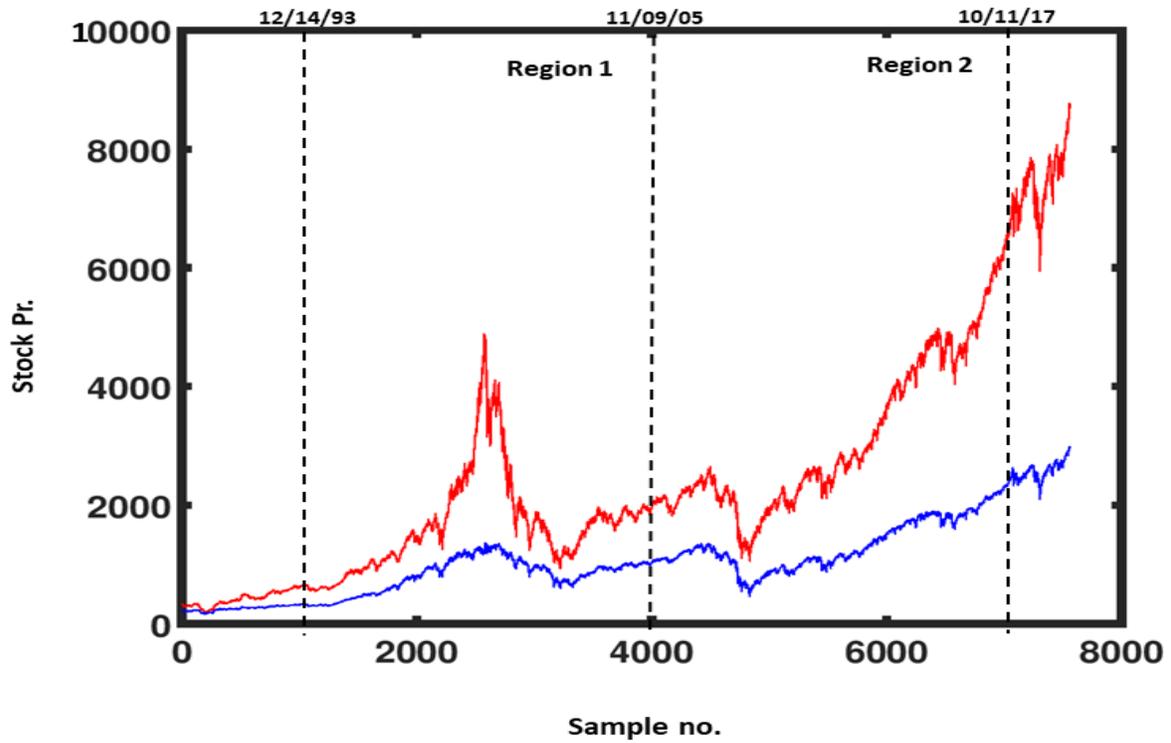

Figure 3. S&P 500 and Nasdaq stock prices for 2 January 1990 – 31 December 2019. Region 1 covers 14 December 1993 – 8 November 2005 and region 2 covers 9 November 2005 – 11 October 2017. Blue – S&P 500. Red – Nasdaq.

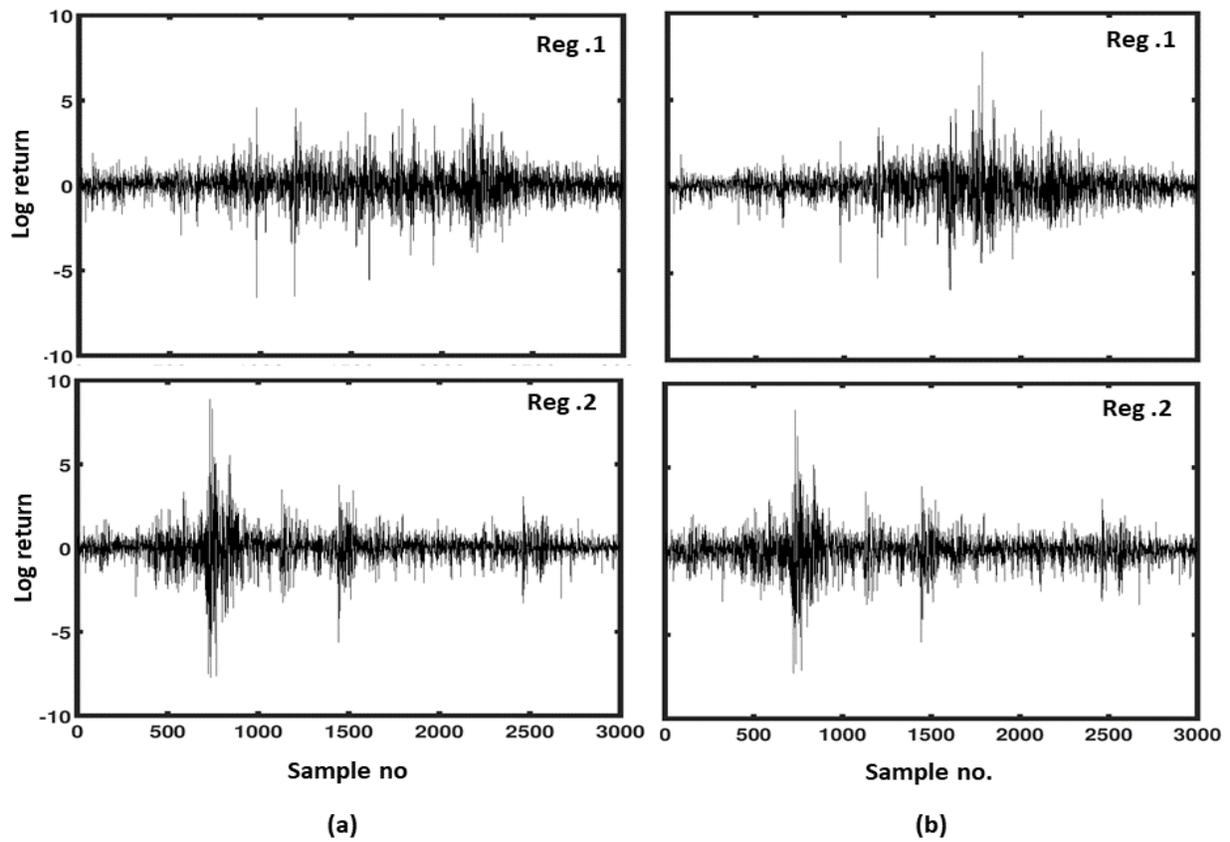

Figure 4. 1 day standardized log returns for region 1 (14 December 1993 – 8 November 2005) and region 2 (9 November 2005 – 11 October 2017) for (a) S&P 500 and (b) Nasdaq.

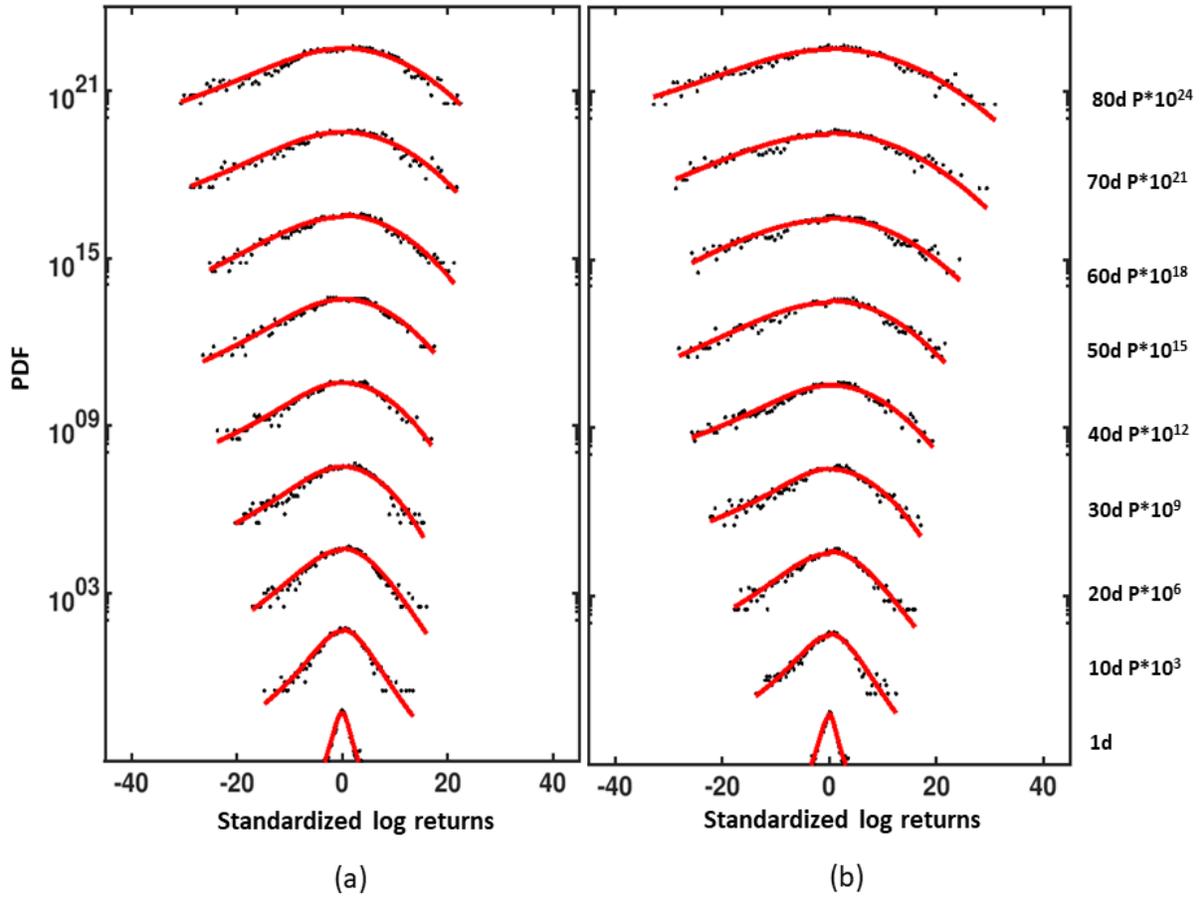

Figure 5. Comparison of the estimated asymmetric $q$-Gaussian distributions with the data distributions for region 1 (14 December 1993 – 8 November 2005). The delays corresponding to the distributions are given on the right hand side of the figure. The distributions for each delay are shifted by multiplying the corresponding PDF with the factors shown on the right hand side next to the delays. (a) S&P 500 and (b) Nasdaq.

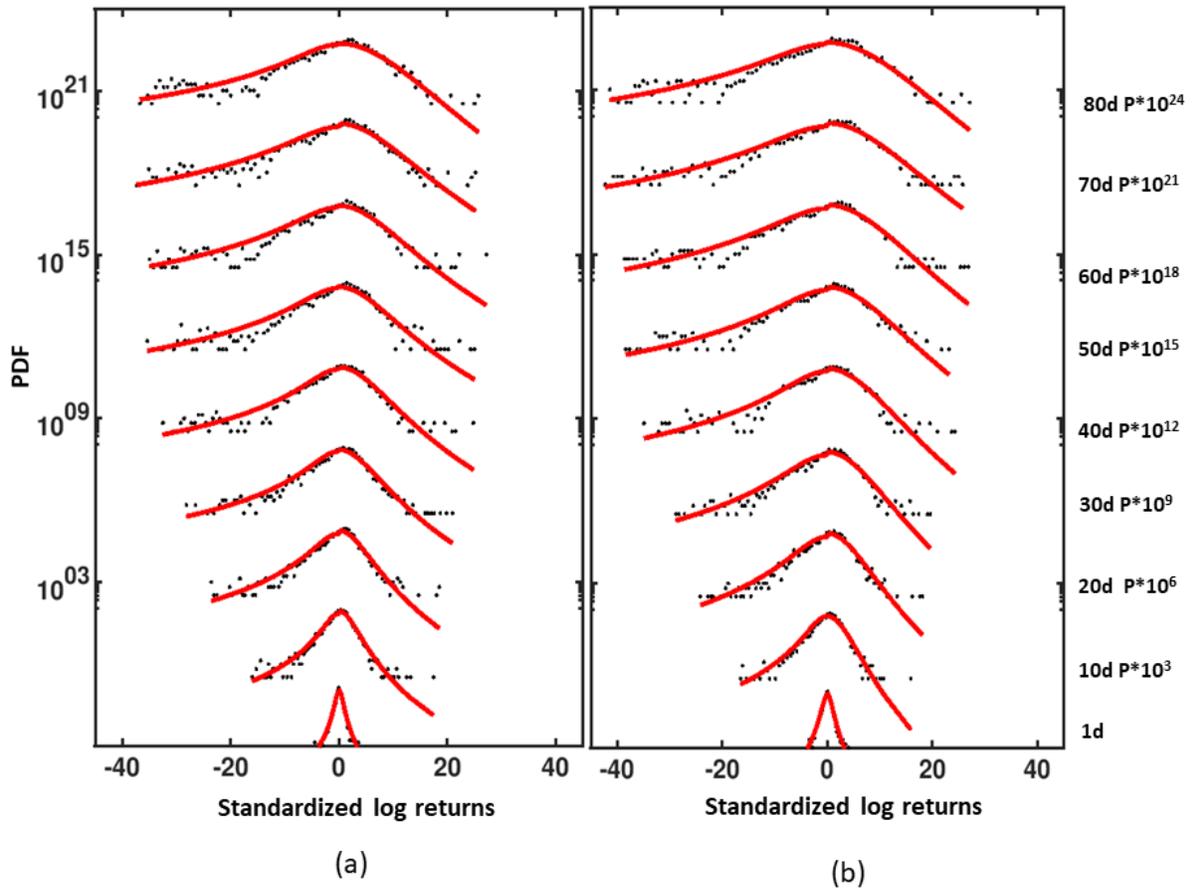

(a)

(b)

Figure 6.  Same as figure 5 for region 2 (9 November 2005 – 11 October 2017).

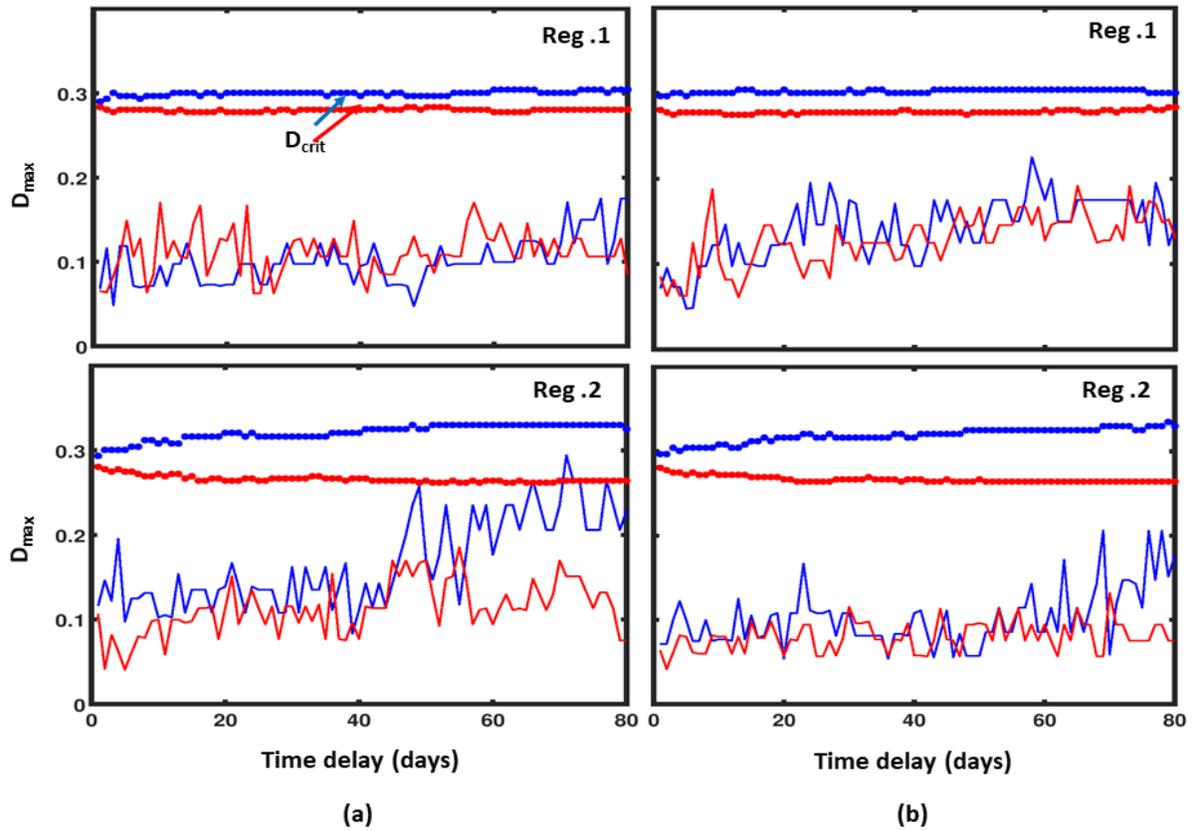

Figure 7. Kolmogorov-Smirnov goodness of fit test for asymmetric $q$-Gaussian distributions. Maximum distance between the empirical and synthetic CDF's are shown as functions of delay for region 1 (14 December 1993 – 8 November 2005) and region 2 (9 November 2005 – 11 October 2017). Also shown in the bold lines are the critical distances for a significance level of 0.05 (confidence 95%). Blue – negative returns, Red - positive returns. (a) S&P 500 and (b) Nasdaq.

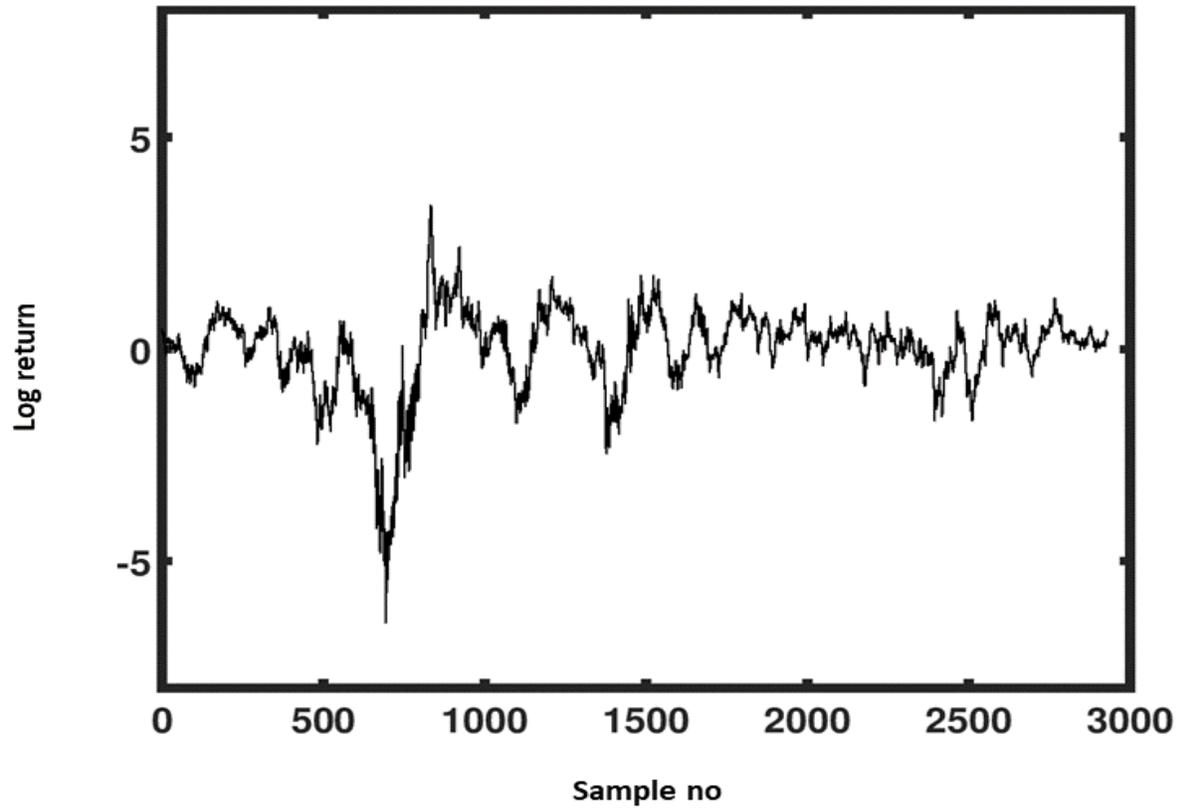

Figure 8.  S&P 500 70 day log return for region 2 (9 November 2005 – 11 October 2017).

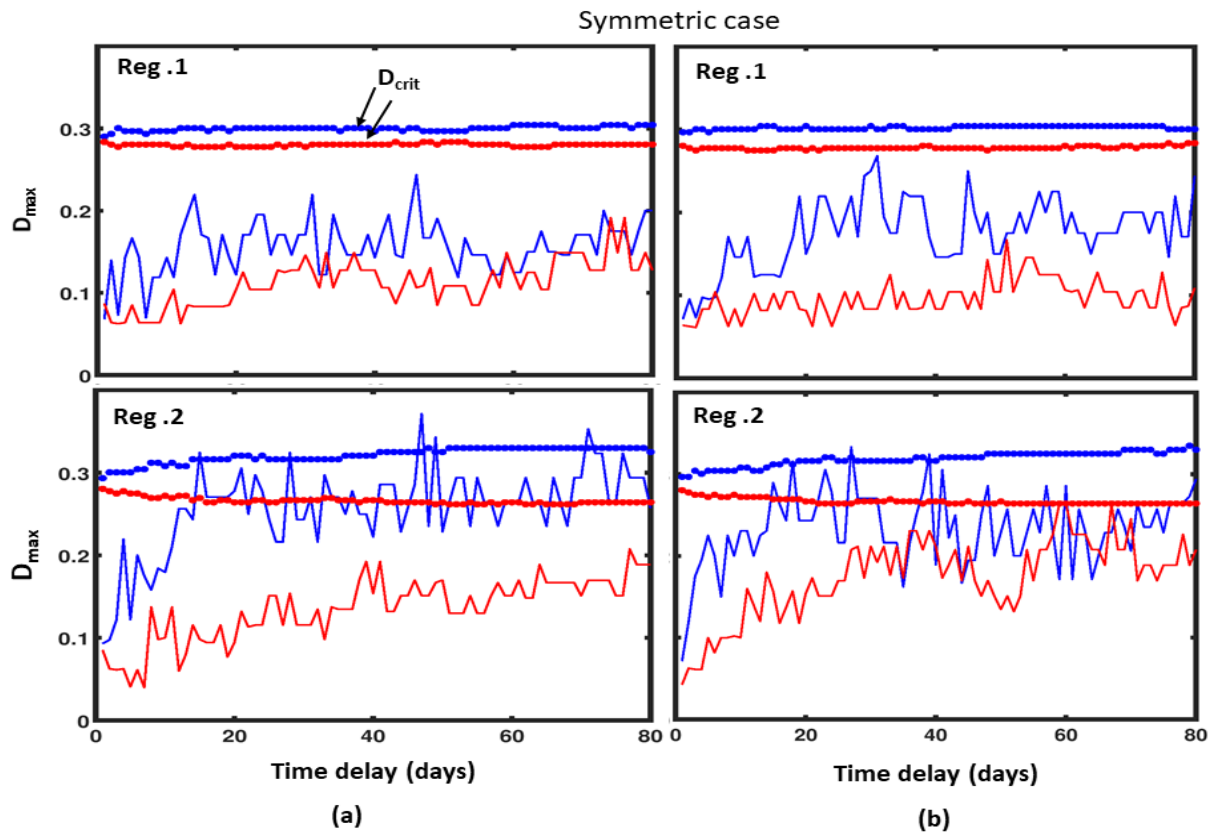

Figure 9. Kolmogorov-Smirnov goodness of fit test for symmetric $q$-Gaussian distributions. Maximum distance between the empirical and synthetic CDF's are shown as functions of delay for region 1 (14 December 1993 – 8 November 2005) and region 2 (9 November 2005 – 11 October 2017). Also shown in the bold lines are the critical distances for a significance level of 0.05 (confidence 95%). Blue – negative returns, Red - positive returns. (a) S&P 500 and (b) Nasdaq.

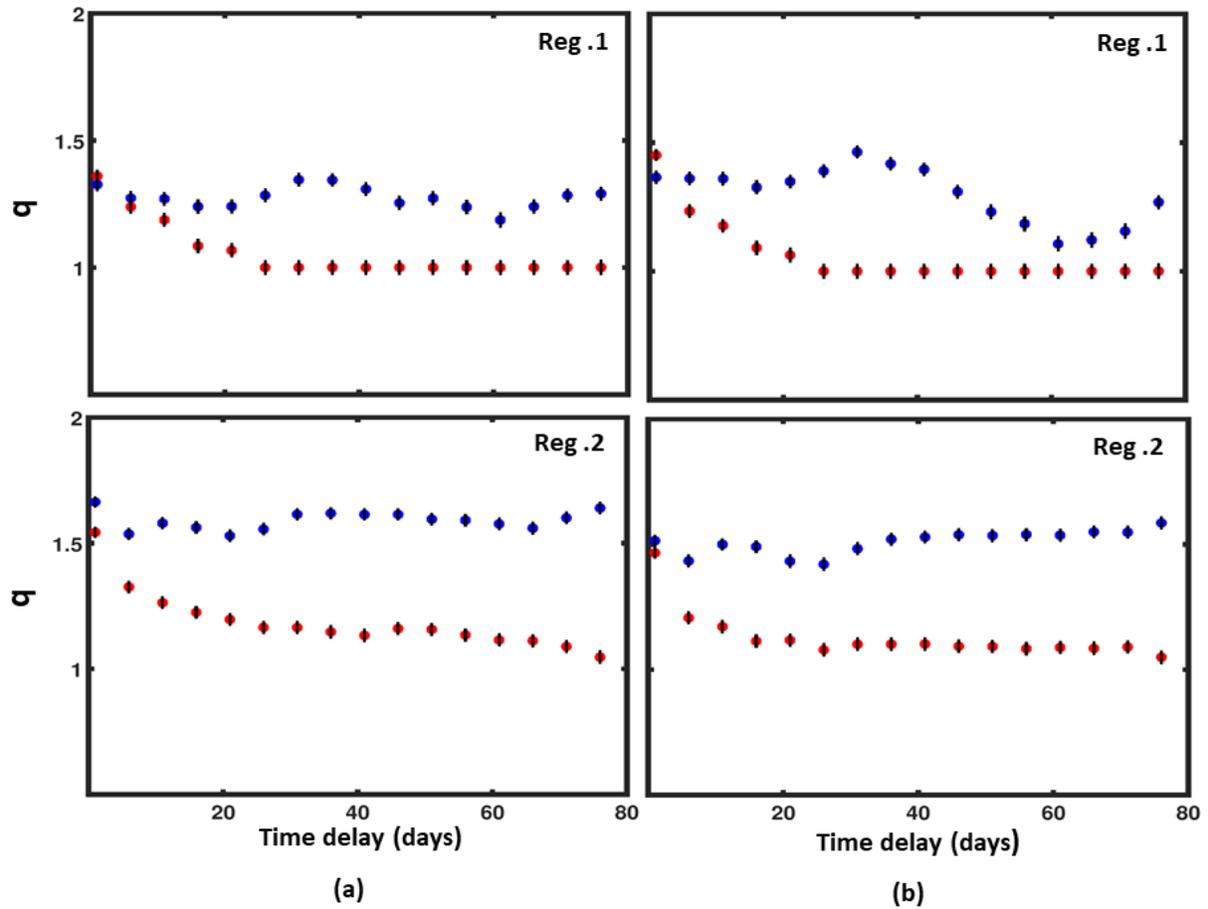

Figure 10. Comparison of $q_-$ (blue) and $q_+$ (red) variation vs. time delay for region 1 (14 December 1993 – 8 November 2005) and region 2 (9 November 2005 – 11 October 2017). Error bars on the estimates are also displayed. (a) S&P 500 and (b) Nasdaq.

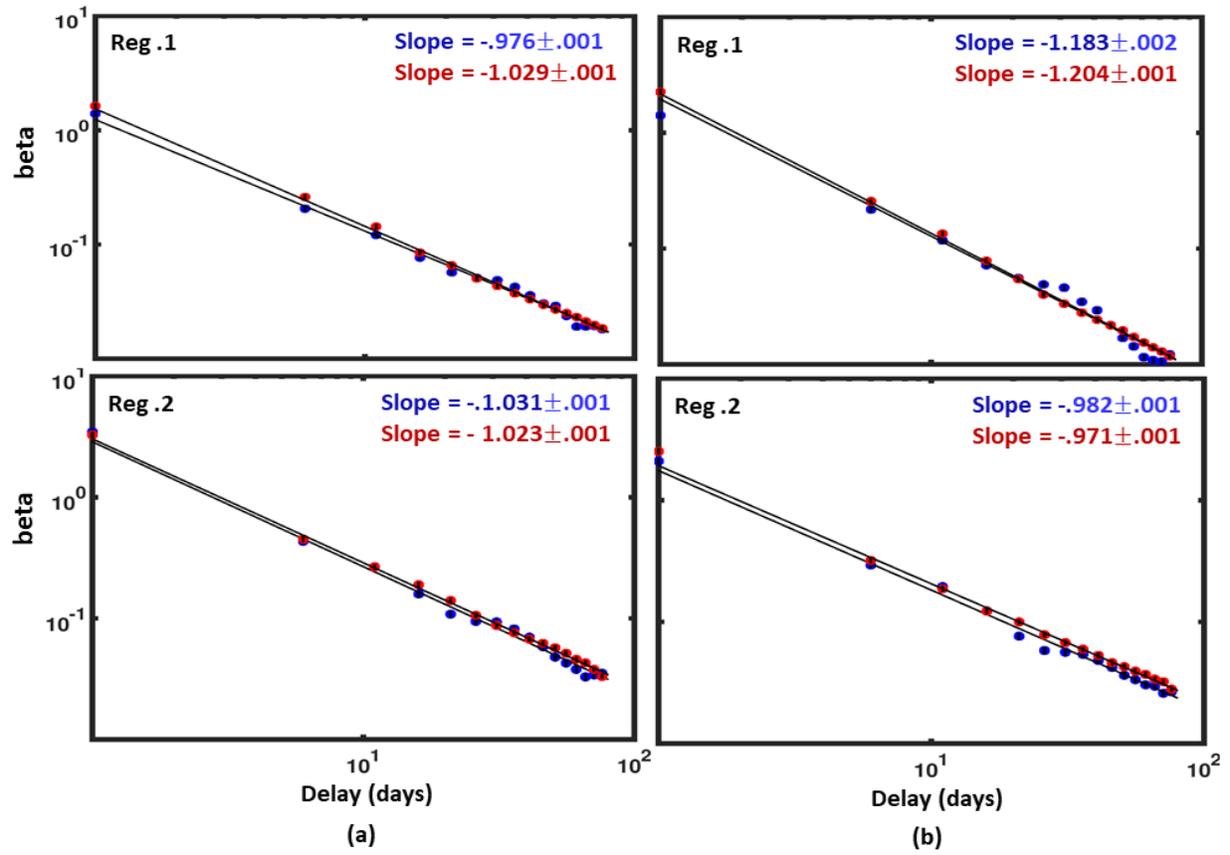

Figure 11. Comparison of the estimated $\beta_-$ (blue) and $\beta_+$ (red) variation vs. time delay for region 1 (14 December 1993 – 8 November 2005) and region 2 (9 November 2005 – 11 October 2017). The error bars are also shown. The solid line is the linear fit. (a) S&P 500 and (b) Nasdaq.

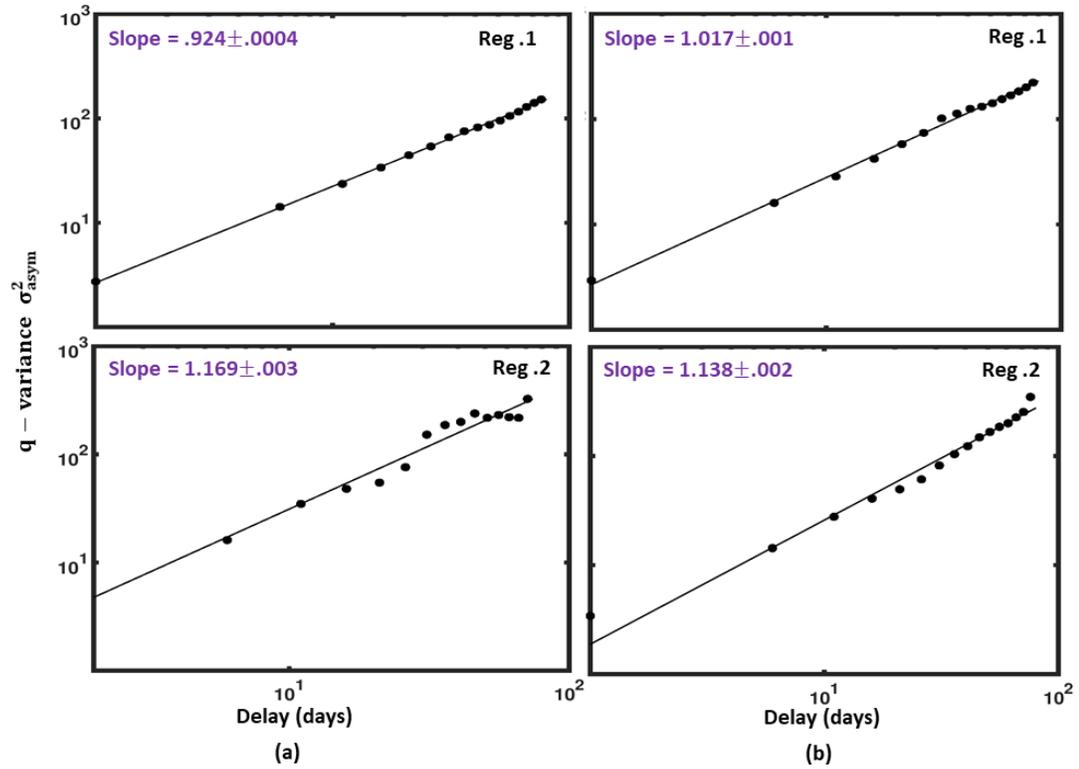

Figure 12. The time scale behavior of $q$-variance $\sigma_{asym}^2$. The $q$-variance is not defined for $q_- > 5/3$ and those are not shown in the figure. The solid line is the linear fit to the log-log data. Region 1 (14 December 1993 – 8 November 2005) and region 2 (9 November 2005 – 11 October 2017). (a) S&P 500 and (b) Nasdaq.

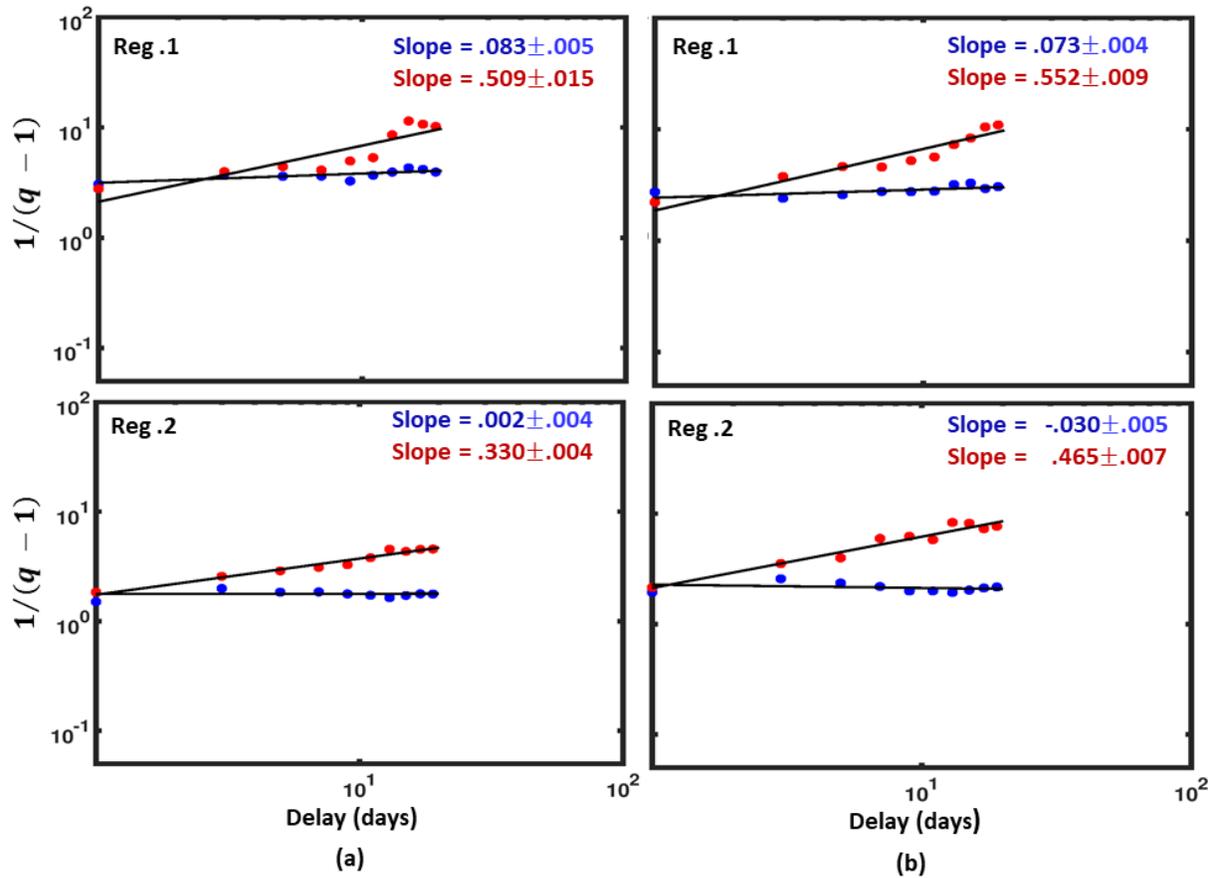

Figure 13. $1/(q-1)$ versus time scale. Blue, negative returns and Red, positive returns. The solid line is the linear fit. Region 1 (14 December 1993 – 8 November 2005) and region 2 (9 November 2005 – 11 October 2017). (a) S&P 500 and (b) Nasdaq.